\documentclass[aps,pra,twocolumn,groupedaddress,floatfix,superscriptaddress]{revtex4-1}

\usepackage{amsmath}
\usepackage{amssymb}
\usepackage{graphicx}
\usepackage{float}
\usepackage[english]{babel}
\usepackage{dsfont}
\usepackage{epstopdf}
\usepackage{soul}
\usepackage{color}
\usepackage{braket}

\usepackage[colorlinks=true,
            linkcolor=magenta,
            urlcolor=blue,
            citecolor=blue]{hyperref}

\usepackage[normalem]{ulem}

\definecolor{Nathanblue}{rgb}{0.12,0.24,0.40}

\def\be{\begin{equation}}
\def\ee{\end{equation}} 
\def\bea{\begin{eqnarray}}
\def\eea{\end{eqnarray}} 
\def\ba{\begin{array}} 
\def\ea{\end{array}}

\def\om{\omega}

\def\nn{\nonumber}
\def\ket{\rangle}

\def\b{\mathbf}
\def\bs{\boldsymbol}
\def\f{\frac}

\def\ra{\rightarrow}
\def\rm{\mathrm}

\newcommand{\mv}[1]{\langle #1\rangle}

\emergencystretch=\maxdimen
\begin{document}

\title{Interaction-induced lattices for bound states: Designing flat bands, \\ quantized pumps and  higher-order topological insulators for doublons}

\author{G. Salerno}
\affiliation{Center for Nonlinear Phenomena and Complex Systems, Universit\'e Libre de Bruxelles, CP 231, Campus Plaine, B-1050 Brussels, Belgium}
\author{G. Palumbo}
\affiliation{Center for Nonlinear Phenomena and Complex Systems, Universit\'e Libre de Bruxelles, CP 231, Campus Plaine, B-1050 Brussels, Belgium}
 \author{N. Goldman}
\affiliation{Center for Nonlinear Phenomena and Complex Systems, Universit\'e Libre de Bruxelles, CP 231, Campus Plaine, B-1050 Brussels, Belgium}
\author{M. Di Liberto}
\email{mar.diliberto@gmail.com}
\affiliation{Center for Nonlinear Phenomena and Complex Systems, Universit\'e Libre de Bruxelles, CP 231, Campus Plaine, B-1050 Brussels, Belgium}

\begin{abstract}

Bound states of two interacting particles moving on a lattice can exhibit remarkable features that are not captured by the underlying single-particle picture. Inspired by this phenomenon, we introduce a novel framework by which genuine interaction-induced geometric and topological effects can be realized in quantum-engineered systems. Our approach builds on the design of effective lattices for the center-of-mass motion of two-body bound states (\emph{doublons}), which can be created through long-range interactions. This general scenario is illustrated on several examples, where flat-band localization, topological pumps and higher-order topological corner modes emerge from genuine interaction effects. Our results pave the way for the exploration of interaction-induced topological effects in a variety of platforms, ranging from ultracold gases to interacting photonic devices.

\end{abstract}

\maketitle

\section{Introduction}

Over the last decades, topology has emerged as a novel paradigm in condensed matter physics. The properties of topological states of matter manifest through various phenomena, including quantized transport coefficients, degeneracies in Bloch bands in the form of Dirac or Weyl points, and the presence of unusual edge modes. In recent years, these features have been identified not only in electronic systems \cite{Kane2010,Qi2011} but also in cold atoms \cite{Goldman2016a, Cooper2019}, photonics \cite{Ozawa_Rev} and many other platforms.

Symmetries and dimensionality play a fundamental role in characterizing the properties of such quantum phases of matter. Topological band theory \cite{Ryu2016} has been developed in order to identify all possible classes of topological phases based on time-reversal, particle-hole and chiral symmetries. This classification has been further extended to include crystalline symmetries \cite{Fu2011,Slager2012}, such as inversion or mirror symmetries, which are related to the point groups of the underlying lattice. More recently, crystal symmetries have been exploited to predict the existence of topological insulators whose edge modes are localized in a lower-dimensional subspace of the geometric edges, such as corners or hinges; these exotic phases have been dubbed \emph{higher-order topological insulators} \cite{Benalcazar_Science, Benalcazar_PRB, Neupert2018}.

Topological band theory relies on a single-particle description (based on quadratic Hamiltonians), where interactions between quasi-particles are absent. Interactions can nonetheless play a crucial role in a variety of topological phenomena \cite{Bergholtz2013, Rachel2018}, as was first revealed in the context of the fractional quantum Hall effect~\cite{Tsui1982, Laughlin1983} and, more recently, in interacting symmetry-protected topological phases~\cite{Haldane1983, Chen2012, Kitaev2010}. And while strongly-correlated many-body systems present some of the most striking phenomena of condensed matter, the physics of few interacting particles can already exhibit highly nontrivial features. For instance, stable (repulsively) bound states of two particles (or \emph{doublons}) can be formed on a lattice thanks to the finite bandwidth of the single-particle dispersion \cite{Mattis1986, Valiente2008}. Tightly-bound doublons typically move through second-order hopping processes similarly to the super-exchange mechanism in Hubbard antiferromagnets. The first observation of repulsively bound states was made possible by the high tunability of cold-atom systems \cite{Winkler2006,Folling2007} and have recently attracted attention for the study of interacting quantum walks \cite{Greiner2015}. Moreover, one-dimensional doublon physics has also been simulated in arrays of linear waveguides \cite{Corrielli2013, Mukherjee2016} and optical fiber setups \cite{Schreiber2012} by exploiting a mapping of the two-body wavefunction into a higher-dimensional free theory \cite{Longhi2011}.

Recently, considerable efforts have been devoted to the study of interaction effects in topological models, and in particular, to the behavior of the resulting two-body states. For instance, in one dimension, onsite interactions were shown to break chiral symmetry and to substantially modify the related edge-state properties. Specifically, studies of the Su-Schrieffer-Heeger (SSH) model have demonstrated how interactions can generate defect potentials on the edge, which can potentially remove topological (bound) edge states or pin non-topological ones \cite{DiLiberto2016, Gorlach2017, Bello2016}; these studies have been extended to the case of nearest-neighbor interactions \cite{DiLiberto2017,Gorlach2017b,Marques2017} and to the case of a mobile impurity \cite{Valiente2019}.  In two-dimensional time-reversal broken systems \cite{Bello2017, Lee2017, Lee2018, Salerno2018}, interactions induce a change of chirality for the doublon edge currents, which have been observed in cold-atom \cite{Greiner2017} and superconducting qubits experiments \cite{Google}. Very recently, topological doublons without a single-particle topological counterpart have been predicted through inhomogenous pair-hopping processes that induce a two-body SSH dimerization, which has been simulated using classical circuits \cite{Gorlach2019}. Besides, the effective mass of bound states in a flat-band system has also been connected to the geometric properties of the single-particle dispersion \cite{Torma2018}. 

In this work, we demonstrate how two-body lattice systems featuring nearest-neighbor interactions, combined with a hardcore constraint, can provide a new framework by which genuine interaction-induced geometric and topological phenomena can be engineered. Our approach builds on the realization of effective lattice structures for doublons: we exploit nearest-neighbor interactions to stabilize a bound state whose center of mass sits on the bonds of the underlying lattice, thus defining a different (dual) lattice for the doublon dynamics. We employ effective theories -- including up to second-order hopping processes -- to describe a variety of intriguing scenarios, which we then validate using numerical exact-diagonalization calculations. First, we show that the simplest configuration featuring isotropic interactions can already yield topological band degeneracies as well as flat Bloch bands for the doublon bound states. In particular, such effective flat bands are associated with a striking interaction-induced localization phenomenon, which can be observed by monitoring the doublon dynamics. Then, we explore regimes where first-order hopping processes dominate and lead to robust topological phenomena for doublons; this includes the realization of interaction-induced Thouless pumps and higher-order topological states, which exhibit unusual edge and corner states of doublons. We finally propose different experimental platforms, such as dipolar gases, Rydberg atoms or topolectrical circuits as realistic systems based on which these interaction-induced phenomena can be investigated. 

\section{Bound states on the dual lattice}

We start by briefly reviewing the theory used below to describe tightly-bound two-particle states on a lattice with nearest-neighbor interactions. Let us consider two hardcore bosons hopping on a lattice according to the Hamiltonian
\be
\label{eq:hub}
H = -\sum_{i\neq j} J_{ij} b^\dag_i b^{}_j  +  \f 1 2 \sum_{\langle i , j \rangle} V_{ij} n_i n_j \equiv H_0 + H_V \,,
\ee
where $H_0$ is the hopping Hamiltonian and we assume that particles interact through nearest-neighbor interactions $H_V$.
Here, we set $V_{ij}= V_{ji} \equiv V_\ell$, where $\ell\equiv(i,j)$ identifies the lattice bonds, and $n_i=b^{\dag}_i b^{}_i$ denotes the number operator. We further impose a hardcore contraint, namely, $b^{}_i b^{}_i = b^{\dag}_i b^{\dag}_i = 0$.

If there exists a set of bonds such that the corresponding interaction $V_\ell$ is larger than all the energy scales of the problem, the set of tightly-bound two-particle states $|d_{\ell}\ket \equiv b^\dag_i b^\dag_j |0\ket$ is the proper basis to describe the problem within the energy window set by $V_\ell$. In the rest of this work, we will refer to such two-body bound states as \emph{doublons}. The center of mass of these composite objects, which sits at the center of the lattice bonds, defines a dual lattice where the doublon dynamics takes place. If a sizeable onsite interaction were present, we would have needed to introduce a second type of doublon state whose center of mass sits on the sites of the lattice. Here, we have removed this case through the hardcore constraint.

\begin{figure}[!t]
\center
\includegraphics[width=1.\columnwidth]{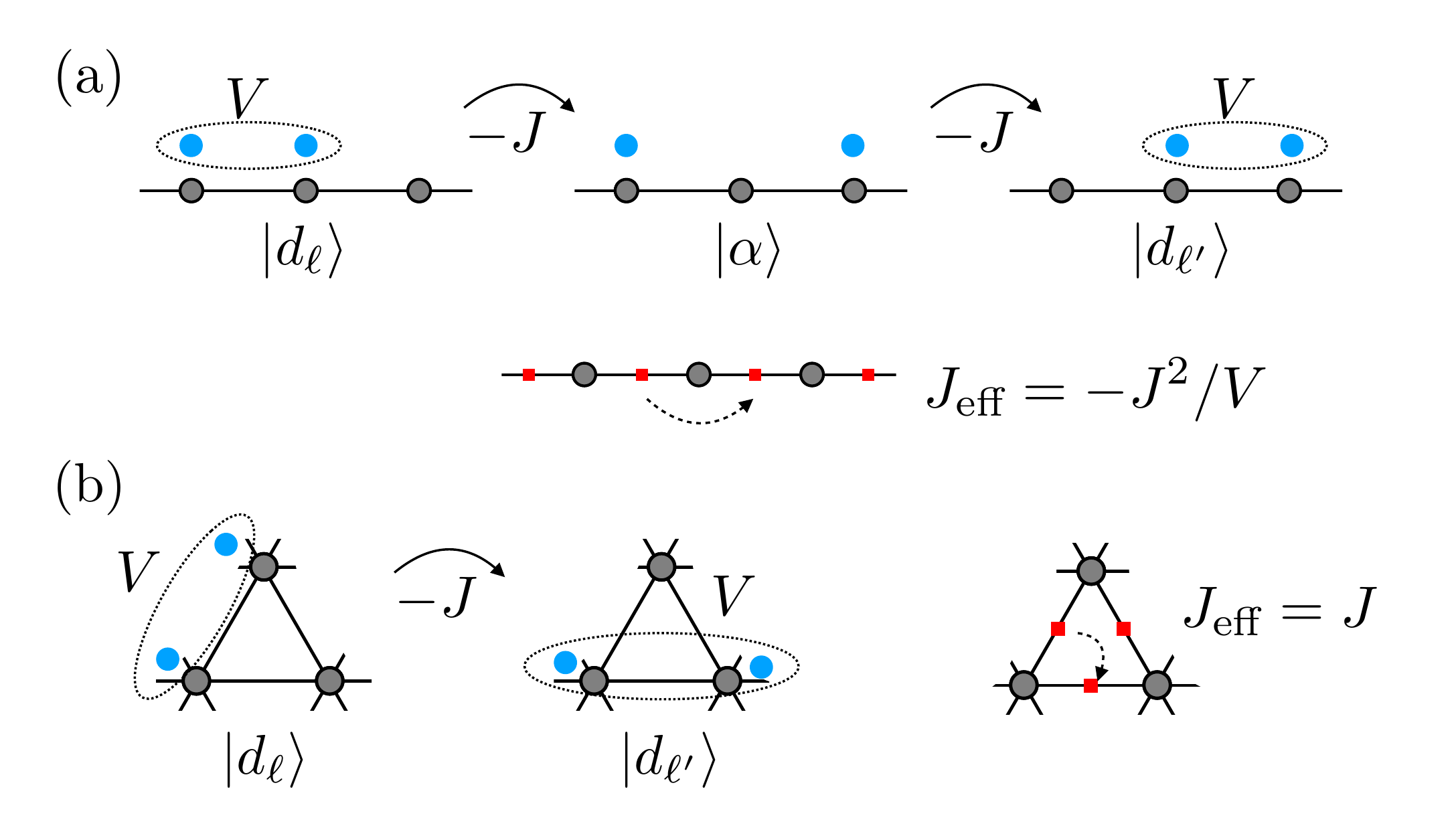}

\caption{Basic doublon dynamics. (a) The first row shows how a bound state $d_\ell$ of energy $V$, with a hardcore constraint, hops through second-order processes on a 1D chain. In the second row, the center-of-mass of the bound state is indicated as square markers corresponding to a dual lattice. The effective doublon hopping amplitude is $J_\textrm{eff}$. (b) First-order hopping process of a doublon in a triangular geometry; the doublon hopping has the same magnitude of the single-particle process.
}
\label{fig:doublon}
\end{figure}

A doublon acquires mobility and it delocalizes across the lattice through single-particle hopping processes that can be described in perturbation theory. Given two doublon states $|d_{\ell}\ket$ and $|d_{\ell'}\ket$, the matrix elements of the effective Hamiltonian $H_{\textrm{eff}}$ describing the doublon Hilbert space can be calculated as
\begin{align}
\label{eq:perturbth}
\mv{d_{\ell} | & H_{\textrm{eff}} |d_{\ell'}} =  V_\ell \delta_{\ell, \ell'} + \mv{d_\ell | H_0 | d_{\ell'}} +  \\
 &+ \f 1 2 \sum_\alpha \mv{d_{\ell} | H_0 | \alpha} \mv{\alpha | H_0 | d_{\ell'}} \left[ \f{1}{V_\ell-\epsilon_\alpha} + \f{1}{V_{\ell'}-\epsilon_{\alpha}} \right] \,, \nn
\end{align}
up to second order in $H_0$  \cite{Cohen}. Here, $|\alpha\ket$ are two-particle eigenstates of $H_V$ whose energies satisfy $\epsilon_\alpha \ll V_\ell$. In our discussion, we neglect the possibility of pair hopping processes but we allow the nearest-neighbor interactions to be anisotropic. When $\ell=\ell'$, the matrix elements correspond to the doublon onsite energies $\epsilon^{\textrm{eff}}_\ell$. When $\ell \neq \ell'$ they correspond to doublon hopping processes that we define as $\mv{d_{\ell} |  H_{\textrm{eff}} |d_{\ell'}} \equiv -J^{\textrm{eff}}_{\ell \ell'}$. The effective doublon Hamiltonian has therefore the following form
\be
\label{eq:eff}
H_{\textrm{eff}} = -\sum_{\ell \neq \ell'} J^{\textrm{eff}}_{\ell \ell'} d^\dag_\ell d^{}_{\ell'}  + \sum_{\ell} \epsilon^{\textrm{eff}}_\ell d^\dag_\ell d^{}_{\ell}\,.
\ee
In Fig.~\ref{fig:doublon}, examples of first-order and second-order processes are shown along with a representation of the center-of-mass position for the doublon bound states. In this framework, the dual doublon lattice can be different from the original one, thus bringing emergent geometric and topological properties. 

In the rest of the paper, we will use the wording \emph{effective model} every time that we show results obtained from the Hamiltonian $H_{\textrm{eff}}$ in Eq.~(\ref{eq:eff}), which describes the center-of-mass motion of the doublon and allows us to treat the interacting problem via a free particle description. In particular, we will use both the real-space and momentum-space representation of Eq.\eqref{eq:eff}. In Appendix \ref{ap:mapping}, we provide an example of the derivation of the effective model Hamiltonian. Instead, when using the wording \emph{two-body model} we refer to the numerical solution of the extended Hubbard model $H=H_0 + H_V$ in Eq.~\eqref{eq:hub} with two particles that we perform by using exact diagonalization in real space. In this case, we will either diagonalize $H$ to obtain the exact spectrum with open boundary conditions or we will use $H$ to compute the time dynamics by calculating the time-evolution operator $\mathcal U = \exp(-i Ht/\hbar)$.

\begin{figure}[!t]
\center
\includegraphics[width=0.95\columnwidth]{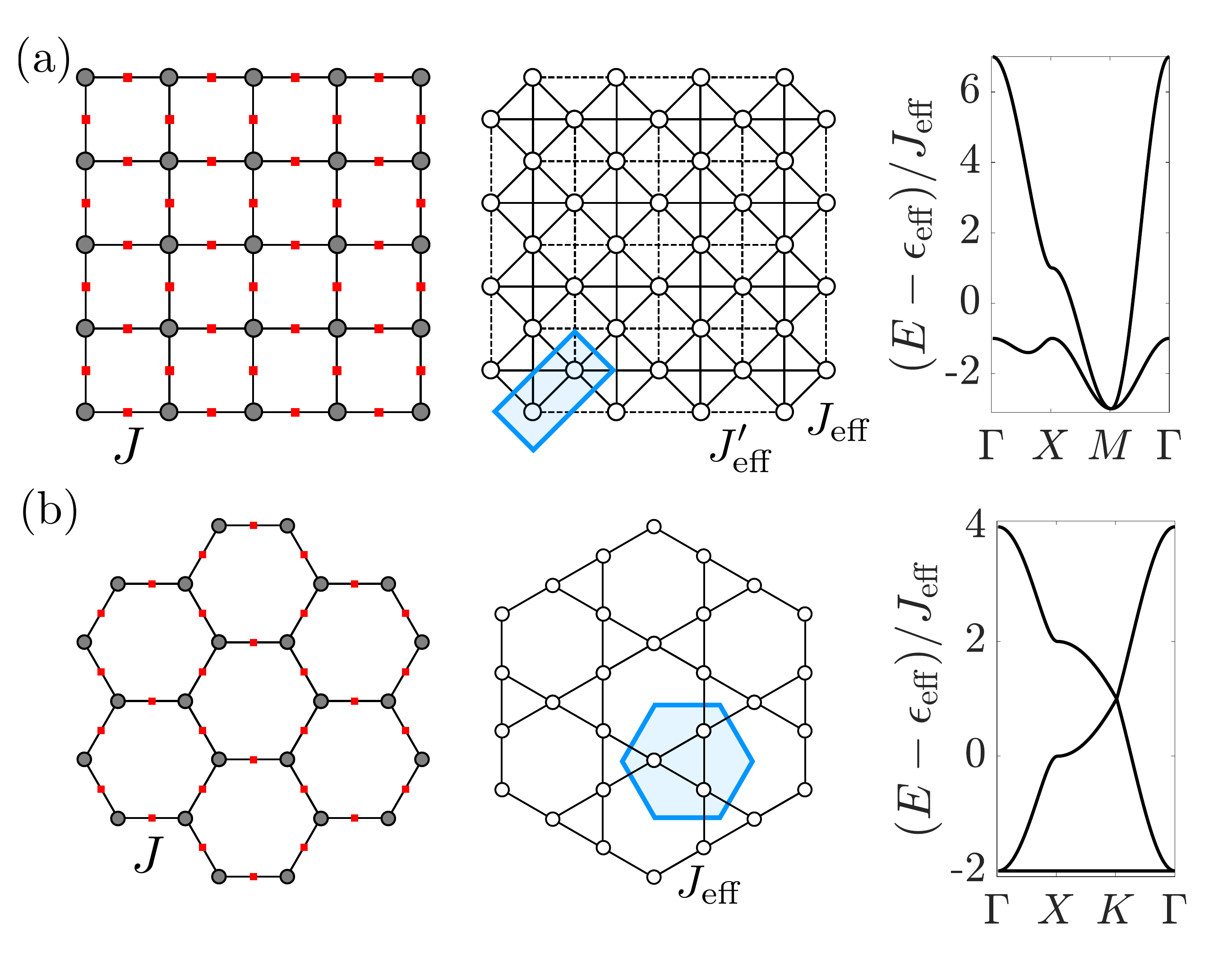}

\caption{(a) The bound state of hardcore bosons on a square lattice with nearest-neighbor interactions is mapped onto a dual chequerboard lattice with hopping coefficients $J'_\textrm{eff} = -J^2/V$, $J_\textrm{eff} = -2J^2/V$ that displays a quadratic band touching minimum at the $M$ point. (b) Starting from a honeycomb lattice, the dual lattice is a kagome lattice with hopping coefficients $J_\textrm{eff}=-J^2/V$ that displays a geometric flat band, a Dirac point at $K$ and a quadratic band touching point at $\Gamma$. The unit cell of the dual lattices is highlighted.
}
\label{fig:squarexmap}
\end{figure}

Let us now consider the simplest case of an interaction with magnitude $V_\ell \!=\! V$ on all nearest-neighbor lattice bonds. On a square lattice, the dynamics of the doublon center of mass takes place on a dual chequerboard geometry, as shown in Fig.~\ref{fig:squarexmap}(a). The emergent dual lattice for the effective model is bipartite and the corresponding energy bands display a topologically protected quadratic band-touching point at $M$ \cite{Sun2009, Montambaux2018} as a function of the center-of-mass momentum. Instead, starting from a honeycomb geometry, the dual lattice is a kagome lattice [Fig.~\ref{fig:squarexmap}(b)]. As a consequence, the doublon dispersion displays an emergent flat band, which induces two-body localization, and also a quadratic band touching point at~$\Gamma$. 

The different scenarios presented so far take place through second-order hopping processes. An interesting perspective occurs when the two-body problem takes place on a lattice that allows for first-order processes in Eq.~(\ref{eq:perturbth}); see Fig.~\ref{fig:doublon} and the following section.

\begin{figure}[!t]
\center
\includegraphics[width=1.\columnwidth]{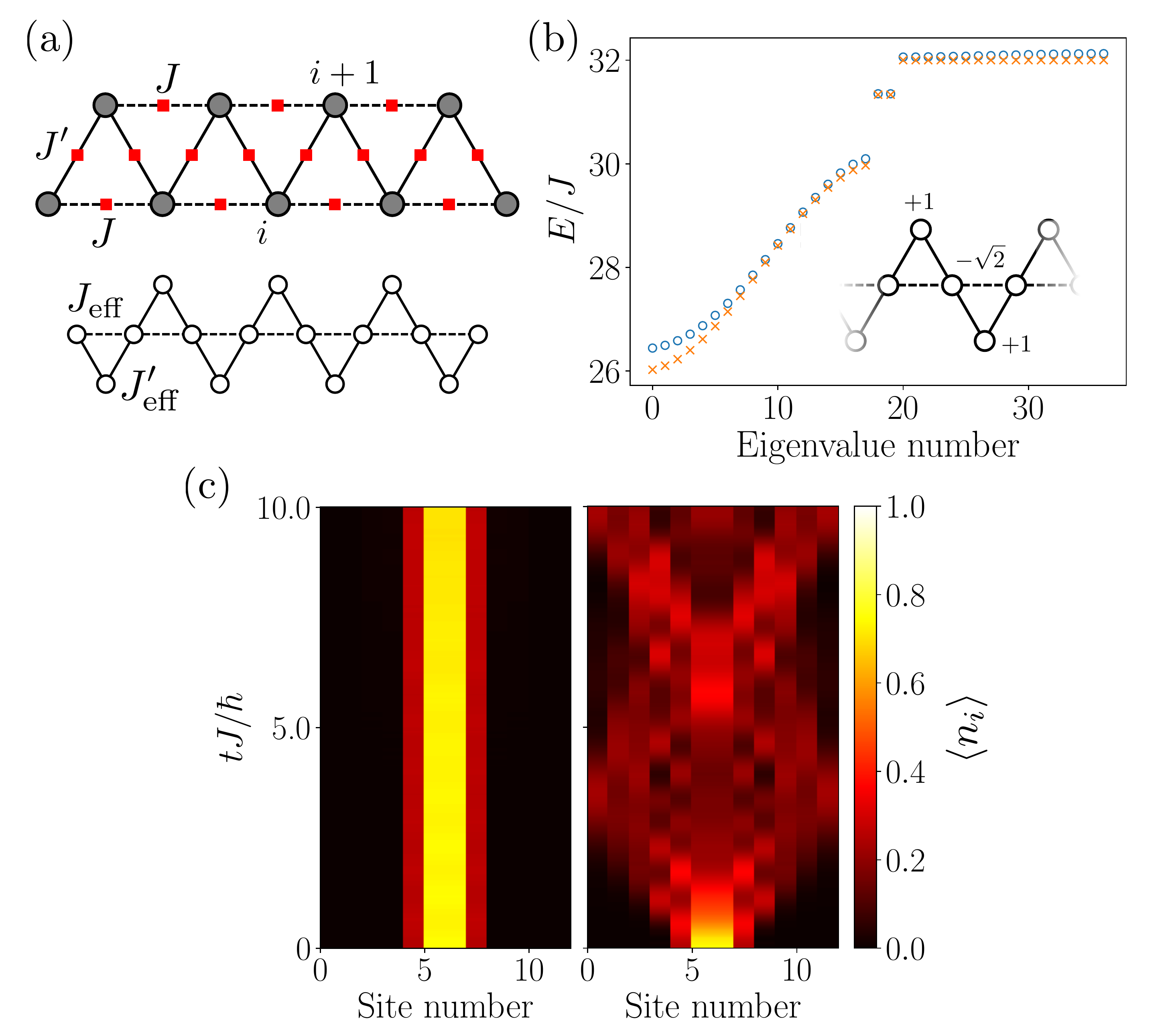}

\caption{ (a) Triangular ladder geometry with uniform nearest-neighbor interactions, and effective dual sawtooth chain geometry. The labels $i$ and $i+1$ indicate the ordering of sites used in the plots and in the text. The single-particle hopping amplitudes are indicated as $J^{\textrm{rung}}= J'$ and $J^{\textrm{legs}}=J$, whereas the effective doublon model described by $J_{\textrm{eff}} = J$ and  $J'_{\textrm{eff}} = J'$ is shown in the second row. (b) Two-body spectrum (circles) and effective model spectrum (crosses) for $V=30J$, $J'=\sqrt{2}J$ and $L=20$ sites, exhibiting a flat band at high energy. The in-gap states are localized edge states that are detuned from the bulk flat band states. (c) Time evolution of particle density $\mv{n_i}$ for the two-body initial state $|\psi_\sigma(0)\ket$ (see main text) for $\sigma = -1$ (left panel) and $\sigma = 1$ (right panel) computed by exact diagonalization; when $\sigma=-1$ the initial state projects onto the flat band, hence yielding flat-band localization for the doublon.
}
\label{fig:triangladdflat}
\end{figure}

\section{The triangular ladder}

\subsection{Isotropic interactions: mapping \\ to the sawtooth chain}

Let us consider a triangular ladder as shown in Fig.~\ref{fig:triangladdflat}(a). If nearest-neighbor interactions $V_\ell = V$ are present for all bonds $\ell$, the dual doublon lattice is equivalent to a sawtooth chain, where the sites have now different connectivities. Let us now take different hopping amplitudes for the rungs and the legs of the ladder, $J^{\textrm{rung}} \!=\! J'$ and  $J^{\textrm{legs}}\!=\!J$. These single-particle hopping processes are inherited by the doublon dynamics on the dual sawtooth chain at first order in Eq.~(\ref{eq:perturbth}). An interesting feature of the sawtooth chain is the presence of a fine-tuned flat band when $J' \!=\! \sqrt 2 J$~\cite{Huber2010}. 

In order to verify the emergent localization properties, we have used exact diagonalization to numerically calculate the two-body spectrum for a chain of $L=20$ sites and $V=30J$, which is shown in Fig.~\ref{fig:triangladdflat}(b). A flat band is clearly visible at high energy, and we note that its very small residual bandwidth is due to hopping processes occurring at second-order or higher in perturbation theory. The localized eigenstates in the flat band of the effective model have the form shown in the inset of Fig.~\ref{fig:triangladdflat}(b), which we use to write the following ansatz for the two-body initial state 
$$
|\psi^\sigma(0)\ket = \left( b^\dag_i b^\dag_{i+1} + \sigma \sqrt 2 b^\dag_{i} b^\dag_{i+2} + b^\dag_{i+1} b^\dag_{i+2} \right) |0\ket
$$ 
in the bulk of the triangular ladder. This two-body state has been time evolved to simulate the corresponding two-body dynamics, as shown in Fig~\ref{fig:triangladdflat}(c) for $\sigma = \mp 1$. In both cases, we have numerically calculated the fidelity of the initial two-particle state $\mathcal F = \sum_\alpha |\mv{ \psi_\alpha | \psi(0) }|^2$, by projecting the initial state onto the set of bound states $|\psi_\alpha\ket$, namely the ones with energy $E_\alpha/J\sim V$. We found that the fidelity is $\mathcal F > 0.991$, thus confirming that the ansatz for the initial state is bound and it will remain bound throughout the dynamics because it has negligible overlap with the scattering two-particle states. 

For $\sigma=-1$, the absence of dynamics indicates that the prepared state well projects onto the effective flat band, which justifies our perturbative approximation truncated to first order; this numerical simulation demonstrates the existence of an interaction-induced localization phenomenon. In contrast, for $\sigma=1$, the initial state substantially projects onto the dispersive band, and it propagates across the lattice while remaining bound, as confirmed by the fidelity with the bound state manifold discussed before, therefore performing a doublon quantum walk. The plot also shows bouncing of the doublon when reaching the two ends of the ladder at $tJ / \hbar \approx 3.5$.

\subsection{Anisotropic interactions: \\ mapping to the SSH chain}

\begin{figure}[!t]
\center
\includegraphics[width=1.\columnwidth]{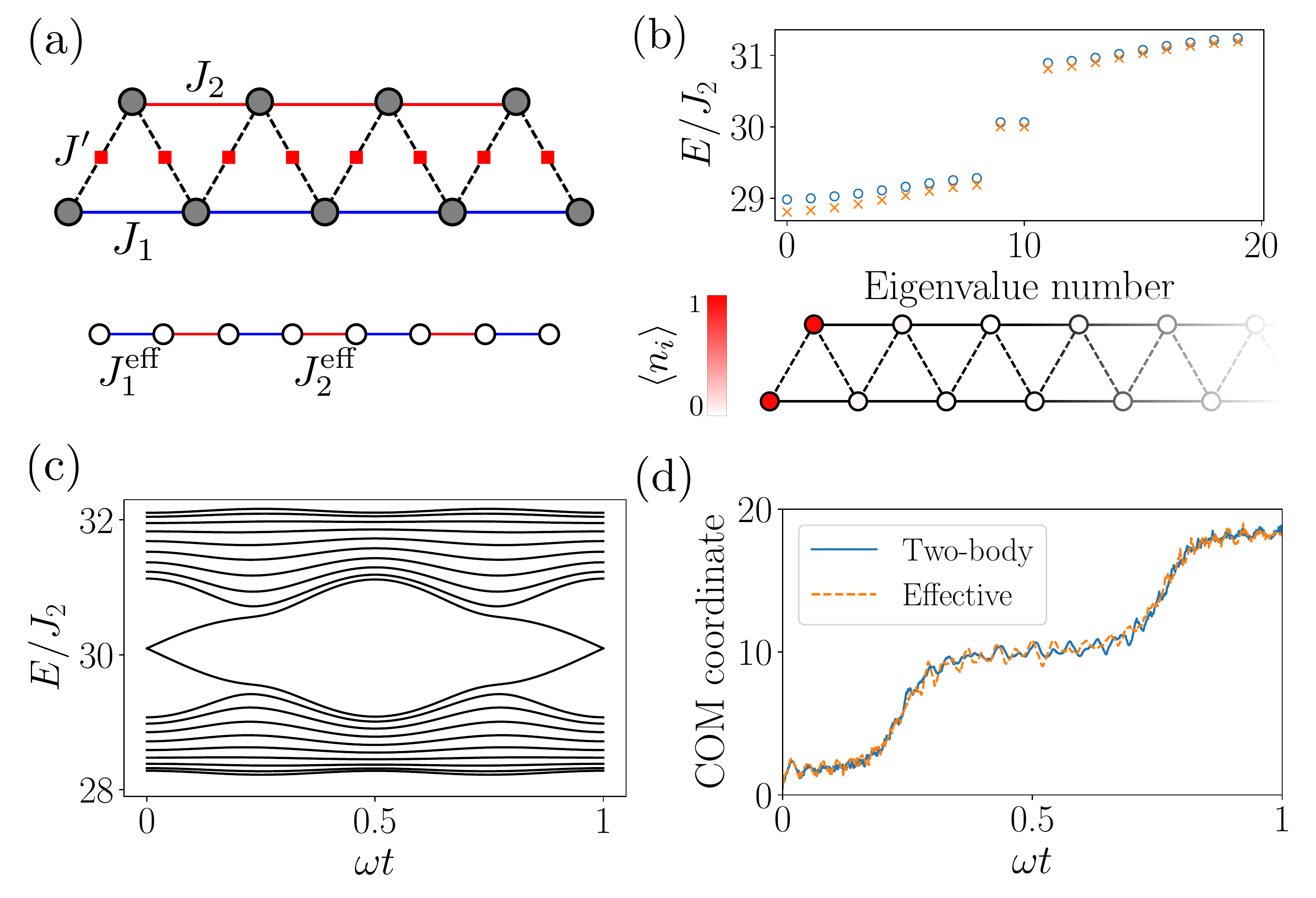}

\caption{(a) Triangular ladder geometry with nearest-neighbor interactions on the rungs, and effective dual SSH geometry. The single-particle hopping amplitudes are indicated as $J^{\textrm{rung}}= J'$ and $J^{\textrm{legs}}=J_{1,2}$, whereas the effective doublon model described by $J^{\textrm{eff}}_{1,2}=J_{1,2}$ is shown in the second row. (b) Two-body spectrum (circles) and effective model spectrum (crosses) for $V=30 J_2$, $J'=J_2$, $J_1=0.2 J_2$ and $L=21$ sites. One of the midgap edge states is shown in the bottom panel. (c) Adiabatic spectrum of the Hamiltonian $H_{\textrm{eff}}$ upon modulation of two-body parameters with $J_0 = 1$, $V = 30 J_0$, $\delta J = 0.5 J_0$ and $\delta V = 0.5 J_0$ (see text). (d) Center-of-mass (COM) dynamics with the parameters as in (c) and $\omega = 10^{-3}J_0$ computed with the (continuous line) two-body and (dashed line) effective Hamiltonians.
}
\label{fig:triangladdssh}
\end{figure}

A novel regime occurs if we instead consider anisotropic interactions. The simplest and most interesting configuration corresponds to having inter-leg interactions only. In this case, the doublon center of mass sits on the rungs of the ladder, and the dual lattice corresponds to a one-dimensional chain, as shown in Fig.~\ref{fig:triangladdssh}(a). By taking different hopping amplitudes on each leg, $J_1,\, J_2$, we obtain a dual dimerized SSH chain for the doublon dynamics. At first-order in perturbation theory, we obtain that the effective doublon hopping amplitudes are $J^{\textrm{eff}}_1=J_1$ and $J^{\textrm{eff}}_2=J_2$, as shown in Fig.~\ref{fig:triangladdssh}(a). As expected, the two-body spectrum shown in Fig.~\ref{fig:triangladdssh}(b) has two gapped bands and admits two-body topological midgap edge states. In Figs.~\ref{fig:triangladdssh}(c) and (d), we show a pumping protocol to transport two-particle states from one edge to the other of the triangular ladder. After preparing the initial state $| \psi (0) \ket = b^\dag_0 b^\dag_{1}|0\ket$ at the left end of the chain, we adiabatically modulate the interactions in time as $V_{2i,2i+1} = V - \delta V \sin(\om t)$, $V_{2i,2i-1} = V + \delta V \sin(\om t)$, where the labeling symbol $i$ follows the same convention as in Fig.~\ref{fig:triangladdflat}. At the same time, we adiabatically modulate the hopping amplitudes on the legs $J_1 = J_0 - \delta J \cos(\om t)$ and $J_2 = J_0 + \delta J \cos(\om t)$. As a result, we observe that the doublon center of mass is dynamically transferred from one edge of the chain to the other one, thus realizing a Thouless pump \cite{Thouless1983, Lohse2015, Takahashi2016} for doublons; see also Ref.~\cite{Greschner2019}.

A comparison with the results of Refs.~\cite{DiLiberto2016,Gorlach2017,Bello2016} is instructive. In these previous works, different aspects of doublon physics have been explored on the SSH chain for large onsite interactions. The resulting effective doublon dynamics was also described by an SSH model, but there the doublon hopping processes only occurred at second order in perturbation theory. Importantly, it was shown that the expected doublon edge modes are missing because interactions induce a strong chiral symmetry breaking at both ends of the chain that is of the same order as the band gap; as a consequence, the doublon edge modes are off-detuned and they become resonant with the doublon bulk bands. In contrast, in our present framework where first-order hopping processes dominate [Fig.~\ref{fig:doublon}(b)], the only source of chiral symmetry breaking comes from higher orders in perturbation theory, which introduce small long-range hopping amplitudes and slightly shift the edge on-site energies compared to the ones of the bulk. These small corrections can be safely neglected for sufficiently large interactions. 

\section{Corner states of doublons in two dimensions}

We now discuss two models in two spatial dimensions, where first order processes [Fig.~\ref{fig:doublon}(b)] provide a mechanism by which corner states of doublons can be generated.

\subsection{Higher-order topological insulator of doublons}

\begin{figure}[!t]
\center
\includegraphics[width=1.\columnwidth]{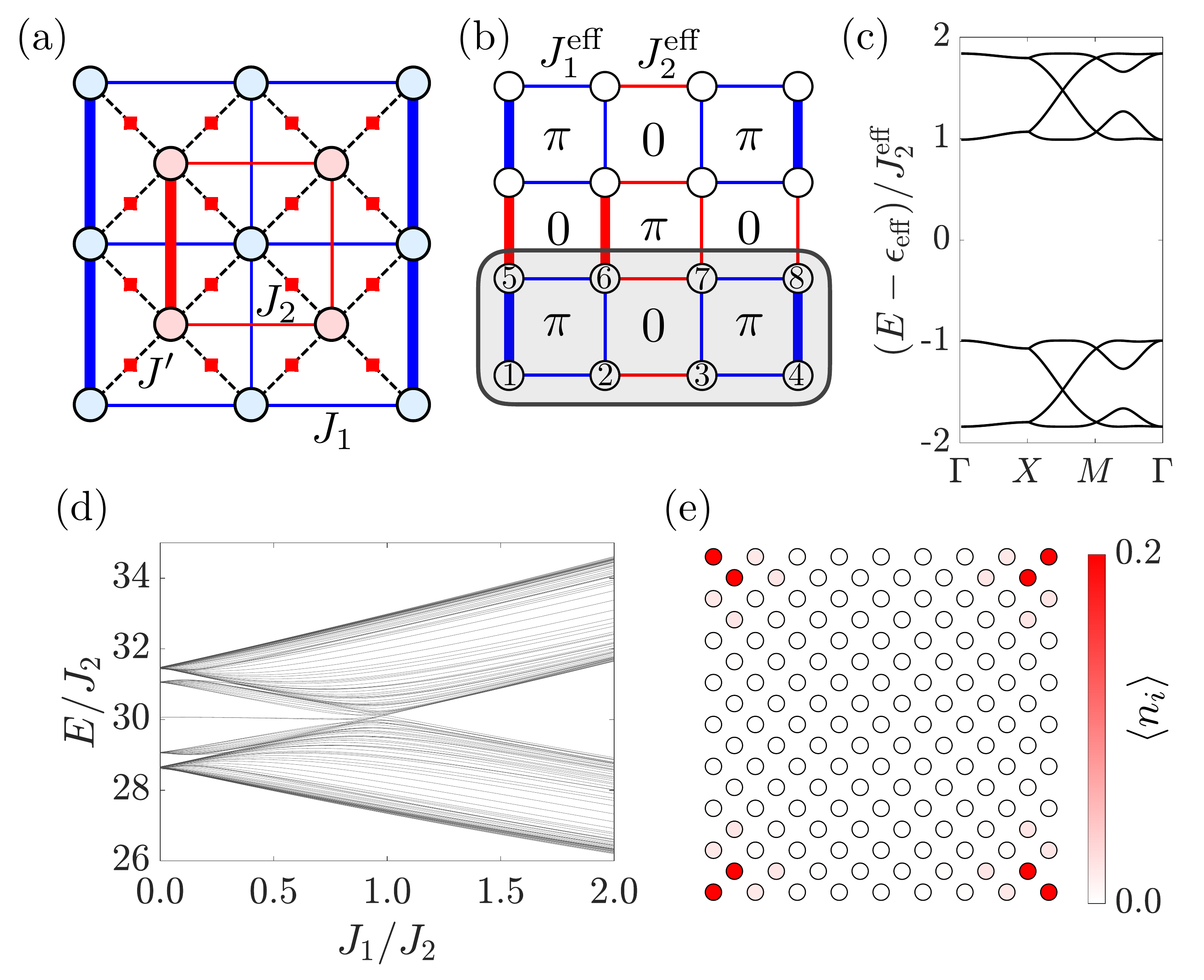}

\caption{(a) Two stacked square lattices (blue and red circles) with different intra-layer hopping amplitudes $J^{\textrm{blue}}\equiv J_1$ and $J^{\textrm{red}}\equiv J_2$ and $\pi$-flux per plaquette in each sublattice (sign changes in the hopping matrix elements are represented by thick vertical lines). Dashed lines represent inter-layer hopping $J'$. (b) Effective model obtained for two hardcore bosons and large nearest-neighbor inter-layer interactions with $J^{\textrm{eff}}_1=J_1$ and $J^{\textrm{eff}}_2=J_2$. The numbers on the lattice sites indicate the ordering used for the effective Hamiltonian in Eq.~\eqref{Ham_hoti}. (c) Band structure of the effective model at first order in perturbation theory for $J_1 = 0.3 J_2$ and $V=30 J_2$. (d) Two-body spectrum as a function of $J_1/J_2$ for $V=30 J_2$. Notice the presence of midgap states for $J_1<J_2$. (e) Doublon corner states density at energy $E\sim 30 J_2$ for $J_1 = 0.3 J_2$ and $J'=0.1 J_2$.
}
\label{fig:hoti}
\end{figure}

Here, we provide an example of an interaction-induced higher-order topological insulator for doublons. The construction of this model is inspired by the triangular ladder in Fig.~\ref{fig:triangladdssh} that we used to generate a SSH doublon model. In this two-dimensional construction, instead of a two-leg ladder we consider two stacked square lattices described by the Hamiltonian
\begin{align}
\label{eq:DLB}
H =& - \sum_{\alpha, m,n} J_\alpha \left( b^{(\alpha) \dag}_{m,n} b^{(\alpha)}_{m+1,n} + b^{(\alpha) \dag}_{2m,n} b^{(\alpha)}_{2m,n+1}  \right)  \\
&-\sum_{\alpha,m,n} J_\alpha e^{i\pi} \, b^{(\alpha) \dag}_{2m+1,n} b^{(\alpha)}_{2m+1,n+1}  \nn\\
&-\f{J'}{2} \sum_{\mv{i,j}}  \left( b^{(1) \dag}_{i} b^{(2)}_{j} + b^{(2) \dag}_{j} b^{(1)}_{i} \right) +\f V 2 \sum_{\mv{i,j}} n^{(1)}_{i} n^{(2)}_{j} \nn \,,
\end{align}
where $\alpha=1,2$ labels the two square lattices and $i=(m,n)$ represents the coordinates of a single lattice point on each lattice. In direct analogy with Fig.~\ref{fig:triangladdssh}, we take two distinct hopping amplitudes on each sublattice or layer, $J_1$ and $J_2$. Moreover, we consider inter-layer interactions $V$, whereas intra-layer interactions are assumed to be negligible. Moreover, we have introduced a magnetic flux $\Phi_B = \pi$ on each lattice by using the same gauge choice as in Ref.~\cite{Benalcazar_Science}, as shown by the Peierls phases in the second line of Eq.~\eqref{eq:DLB}. Since the intra-layer hopping coefficients are uniform, each separate lattice corresponds to the gapless limit of the higher-order topological insulator model, known as Benalcazar-Bernevig-Hughes (BBH) model, introduced in Ref.~\cite{Benalcazar_Science}. A graphical representation of the model in Eq.~\eqref{eq:DLB} is shown in Fig.~\ref{fig:hoti}(a). The stacking is chosen such that one lattice sits on the plaquette centers of the other. 

The dual doublon lattice is shown in Fig.~\ref{fig:hoti}(b), and it displays an emergent staggered flux pattern. We show now that this model shares several properties with the BBH model introduced in Ref.~\cite{Benalcazar_Science}, which is the prototypical example of a higher-order topological insulator. Differently from that case, the unit cell here includes $8\times8$ sites, as shown in Fig.~\ref{fig:hoti}(b) where the unit cell ordering that we choose is indicated. After Fourier transforming the effective model Hamiltonian, we obtain 
\be
H^{\textrm{eff}} (\mathbf{k}) =
\begin{pmatrix}
\Sigma & \Delta \\
\Delta^* & \Sigma
\end{pmatrix}\label{Ham_hoti}
\ee
with
\be
\Sigma = 
\begin{pmatrix}
J_1 \sigma_x & c_1\sigma_x + c_2\sigma_y \\
c_1^*\sigma_x + c_2^*\sigma_y & J_1 \sigma_x
\end{pmatrix}
\ee 
and $\Delta = \sigma_z \otimes (\tilde{c}_1 \sigma_0 + \tilde{c}_2 \sigma_z)$, where $c_1 = J_2(1+e^{-ik_x})/2$, $c_2=J_2(1-e^{-ik_x})/2i$, $\tilde{c}_1=-J_2e^{-ik_y}$, $\tilde{c}_2=-J_1$, $\sigma_{x,y,z}$ are the Pauli matrices and $\sigma_0$ is the $2\times2$ identity matrix. The band structure of the effective model is shown in Fig.~\ref{fig:hoti}(c).

In the limit where $J_1 = 0$, this model is equivalent to the topologically non-trivial atomic limit of the BBH model up to a basis transformation. Therefore, it displays a bulk quadrupole moment and corner states protected by two non-commuting mirror symmetries. In our basis, the mirror symmetries have the matrix form $M_x = \sigma_z\otimes\sigma_y\otimes\sigma_y$ and $M_y = \sigma_x\otimes\sigma_0\otimes\sigma_0$. Additionally, this model always has a chiral symmetry $\mathcal{C} = \sigma_z\otimes\sigma_0\otimes\sigma_z$. For finite $J_1$, only $M_y$ and $\mathcal{C}$ remain symmetries of the model and the quantization of the bulk quadrupole moment is therefore not guaranteed; however, the chiral symmetry still protects the corner states \cite{Brouwer2017}, which are expected to remain at the mid-gap energy. Instead, in the limit where $J_2 = 0$ the model corresponds to the topologically trivial atomic limit of the BBH model, which has no quadrupole moment and no corner states.

In Fig.~\ref{fig:hoti}(d), we show the exact two-body spectrum of this model with open boundary conditions as a function of the ratio $J_1/J_2$. For $J_1<J_2$, we find four corner modes at mid-gap energy that exist until the gap closes, namely for $J_1=J_2$. An example of the two-body corner modes is shown in Fig.~\ref{fig:hoti}(e). A topological transition takes place at $J_1 = J_2$ upon closing the band gap, which is supported by the disappearance of the corner modes. 

\subsection{Corner modes of doublons on the Lieb lattice}

\begin{figure}[!t]
\center
\includegraphics[width=1.\columnwidth]{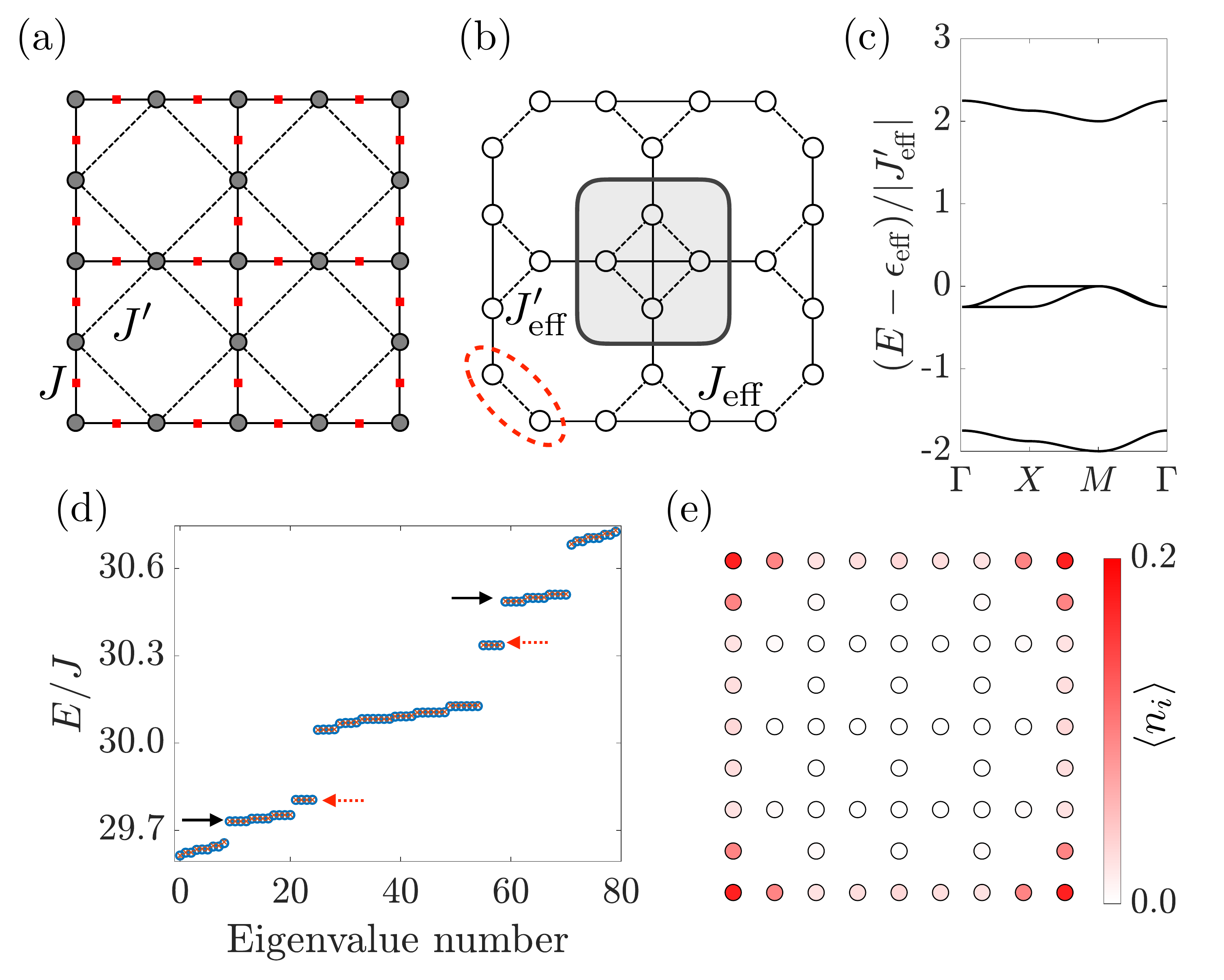}

\caption{(a) Lieb lattice with nearest-neighbor hopping $J$ and next-nearest-neighbor hopping $J'$ with interaction $V$ on the latter bonds. (b) Square-octagon dual lattice with effective hopping amplitudes $J'_{\textrm{eff}}=J'-J^2/V$ and $J_{\textrm{eff}}=-J^2/V$. The corners of this effective model are highlighed by a red dashed line and correspond to dimers. (c) Doublon band structure for the effective square-octagon model for $J'=0.3 J$ and $V=30J$. (d) Two-body spectrum with open boundary conditions for a Lieb lattice with $4\times 4$ plaquettes and parameters as in (c). Corner modes are indicated by red dashed arrows. Notice the presence of additional bands indicated by solid black arrows corresponding to edge states. (e) Doublon corner states density for the upper energy gap (d).}
\label{fig:squareoct}
\end{figure}

As a final example, we consider a Lieb lattice (or a decorated square lattice) with nearest-neighbor interactions $V$, nearest-neighbor hopping $J$ and next-nearest-neighbor hopping $J'$, as shown in Fig.~\ref{fig:squareoct}(a). The dual lattice for the doublon center of mass is a square-octagon lattice [see Fig.~\ref{fig:squareoct}(b)], and the bound state dynamics is governed by a combination of first-order and second-order processes that yield two hopping amplitudes,  $J'_{\textrm{eff}}=J'-J^2/V$ and $J_{\textrm{eff}}=J^2/V$. The corresponding doublon band structure is displayed in Fig.~\ref{fig:squareoct}(c). We point out that several studies have investigated the properties of this lattice in the presence of magnetic flux \cite{Pal2018} or spin-orbit coupling \cite{Fiete2010}, as well as in Kitaev-type models \cite{Kells2011, Hermanns2017}.

An interesting feature of the dual lattice emerges upon applying open boundary conditions: at the corners of the lattice, the unit cell breaks into dimers, as illustrated in Fig.~\ref{fig:squareoct}(b). In the atomic limit where $J_{\textrm{eff}}\ra 0$, namely when $V\ra\infty$, corner modes sit on a dimer and have energy $\epsilon_{\textrm{corner}} \!=\! V \pm J'$ whereas the bulk bands are centered around $\epsilon^\pm_{\textrm{bulk}} \!=\! V\pm 2 J'$ and  $\epsilon^0_{\textrm{bulk}} \!=\!V$. As a consequence, the corner modes are off-resonant with respect to the bulk modes and they can be spectrally resolved. These corner modes are not protected by the two commuting mirror symmetries of the model, however for finite values of $J_{\textrm{eff}}$ they remain in the band gap, as shown in Figs.~\ref{fig:squareoct}(d)-(e). Differently from intrinsic corner modes, these states can be seen as termination-dependent (or \emph{extrinsic}) corner states \cite{Brouwer2017,Palumbo2016}; such edge states occur in other lattices, such as the sawtooth chain (see Fig.~\ref{fig:triangladdflat}) or in the diamond chain with $\pi$-flux \cite{Mukherjee2018}. Similar arguments hold for the edges of the lattice, where the unit cell breaks into a set of three sites, thus yielding the localized edge modes shown in the spectrum of Fig.~\ref{fig:squareoct}(d).

\section{Discussion and Conclusions}

In this work, we have discussed how nearest-neighbor interactions dramatically affect the physics of two-body bound states on different lattice systems. Moreover, the presence of a hardcore constraint plays a crucial role in setting a well defined dual lattice for the dynamics of the bound-state's center of mass. While our calculations have been performed for hardcore bosons, spin polarized fermions could also be considered.

In recent years, several tuneable platforms with long-range interactions have become available. Dipolar long-range interactions appear in Rydberg atoms, which have been recently exploited to investigate a many-body symmetry-protected topological phase \cite{Browaeys2019}, or to create bound atomic dimers at distances comparable to the lattice spacing \cite{Gross2019}. In dipolar gases \cite{Pfau2009}, long-range interactions have allowed for the experimental exploration of the Mott insulator-superfluid transition in an extended Bose-Hubbard setting \cite{Ferlaino2016}. 
 
A concrete platform could be offered by dipolar atoms in optical lattices, where similar preparation and probing methods as those used in the two-body experiment~\cite{Greiner2017} could be considered. The configurations depicted in Figs.~\ref{fig:squarexmap}, \ref{fig:triangladdflat} and \ref{fig:squareoct} would require a polarization axis orthogonal to the lattice plane such that the dipolar interaction acts isotropically on all the nearest-neighbor bonds. The control over the relative single-particle hopping strengths, $J$ and $J'$, required in the model illustrated in Fig.~\ref{fig:triangladdflat} can be achieved by shaking the optical lattice along the ladder axis. In order to access the physics described in Fig.~\ref{fig:triangladdssh}, one would need to consider a squeezed triangular ladder such that the distance between neighboring sites on each leg of the ladder is larger than the distance between neighboring sites belonging to different legs. The power-law decay of the dipolar interactions will then provide a dominant contribution to the shortest bonds. The pumping scheme described in Sec.~III B requires a control over the hopping amplitudes $J_1$ and $J_2$, which could be achieved by using an anisotropic optical lattice (whose depth is different on different legs of the ladder). Finally, the staggered inter-leg interaction can be obtained by exploiting the anisotropic nature of dipolar interactions, namely by appropriately tilting the polarization axis, e.g. along the plane orthogonal to the even lattice bonds. The model described in Fig.~\ref{fig:hoti} could be instead simulated by using state dependent lattices in a similar fashion as proposed in Ref.~\cite{Lee2019}. One species would experience a lattice positioned on the blue sites with hopping $J_1$ whereas the other species would experience a different lattice positions on the red sites with hopping $J_2$. Inter-species dipolar interactions would dominate on the shortest lattice bonds, namely the inter-lattice bonds, thus binding the two types of atoms. The additional difficulty would be to realize the $\pi$ flux per plaquette on each lattice, which requires synthetic gauge field methods \cite{Goldman2016a, Cooper2019}.

Another route to realize the results presented in this work could be to exploit a mapping of the two-body wavefunction into a higher dimensional but single-particle theory \cite{Longhi2011, Schreiber2012, Corrielli2013, Mukherjee2016, DiLiberto2016, Gorlach2017}. In this case, it would be possible to exploit a linear platform to build a simulator of two-particle lattice models. While one-dimensional chains with two-interacting particles have been already simulated by using two-dimensional arrays of optical waveguides or optical fiber networks, interacting models with more involved lattice connectivity or dimensionality higher than one seem quite unrealistic. To this purpose, a better advantage could be provided by topolectrical circuits \cite{Simon2015, Liang2015, Thomale2018}, as demonstrated by a recent experiment simulating topological two-body physics \cite{Gorlach2019}. In this platform, lattice connectivity is highly flexible and recent advances have shown that even four-dimensional topological models can be realized \cite{Price2020}. This latter case is quite relevant because the models discussed in Sec.~IV would require a mapping to a four-dimensional free theory.

The results presented in this work pave the way to novel and interesting questions concerning bound-states physics. In our work, we have investigated scenarios in quasi-1D and 2D geometries, but the physics of 3D topological doublons is mostly unexplored. An interesting perspective is to understand the effects of dipolar long-range interactions, which will manifest their anisotropic nature together with a competition between attractive and repulsive behavior. Moreover, while the effects of magnetic flux and nearest-neighbor interactions on doublon physics has only been considered for the Harper-Hofstadter model \cite{Lee2017}, a rich behavior can appear for the geometries considered in our work or in 3D, where more exotic regimes, e.g. hosting chiral hinge states \cite{Benalcazar_PRB}, can exist. 

As a concluding remark, it is worth calling the attention on the fact that doublon systems can display nontrivial many-body phases upon appropriate quench protocols \cite{Petrosyan2007,Rosch2008,Kantian2010}. At the same time, recent quantum distillation experiments have demonstrated the possibility to purify cold atomic gases in optical lattices from unbound states (singlons) thanks to the faster mobility of the latter \cite{Weiss2015, Aidelsburger2018}. An important outlook is therefore to understand the quantum distillation mechanism in the framework considered in our work. Indeed, even though we have argued that first-order single-particle processes can be dominant for the doublon dynamics, their mobility is nevertheless affected by the effective dispersion of the dual lattice as compared to the one of unbound particles. This fact could be therefore exploited to investigate  quantum doublon distillation in the many-body regime with long-range interactions.

Finally, we note that recent advances in photonics~\cite{Google, Simon2019} have allowed for the creation of strong on-site interactions for photons, which opens an interesting avenue for two-body topological physics in that context.

While preparing this manuscript, we became aware of a recent work \cite{Lee2019} investigating topological bound states on a triangular ladder with nearest-neighbor interactions, as realized in state-dependent lattices. Besides, another recent study \cite{Grusdt2019} has shown how fractional corner charges appear in the ground state of a 2D SSH lattice with Bose-Hubbard interactions.

\section{Acknowledgements}

The authors would like to acknowledge I. Carusotto and L. Barbiero for insightful discussions. This work was supported by the ERC Starting Grant TopoCold, and the Fonds De La Recherche Scientifique (FRS-FNRS, Belgium).


\appendix

\section{Example of derivation of the effective model for the honeycomb lattice}
\label{ap:mapping}

In this Appendix, we provide an example that shows how effective models for two-body bound states can be constructed. A similar reasoning is also presented in Refs.~\cite{DiLiberto2016, Salerno2018} for onsite interactions. 

\begin{figure}[!t]
\center
\includegraphics[width=1\columnwidth]{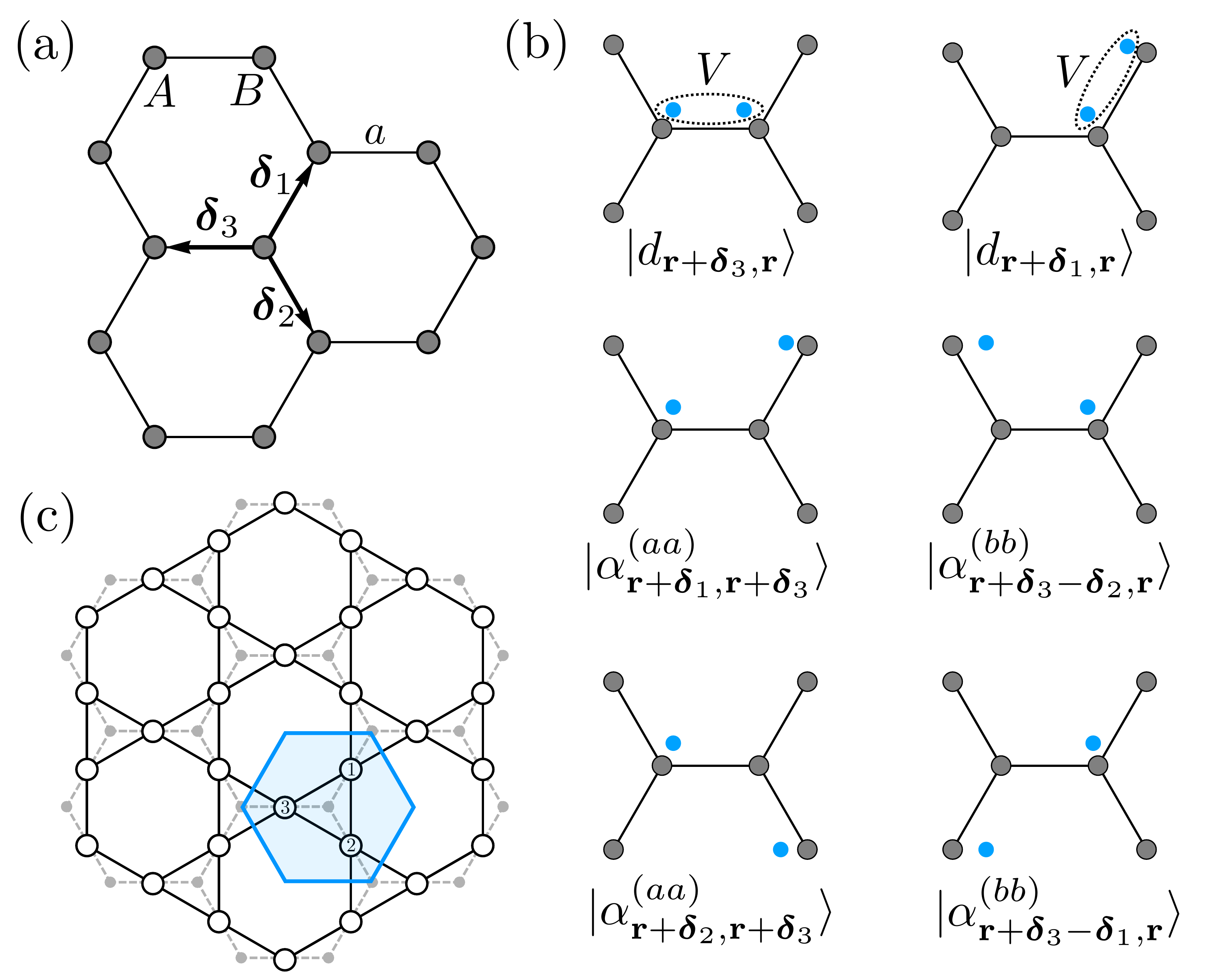}
\caption{(a) Honeycomb lattice geometry, where $A$ and $B$ label the two inequivalent sites of the unit cell. The vectors $\bs \delta_{1,2,3}$ connect nearest-neighbor sites. 
(b) Visual representation of the doublon and the virtual states used to obtain Eqs.~\eqref{appendix_e_eff} and~\eqref{appendix_J_eff}.
(c) Dual Kagome lattice geometry for the doublon center of mass, where the unit cell is also highlighted. The honeycomb lattice is also shown. The numbers on the lattice sites inside the unit cell indicate the ordering used for matrix representation of the effective Hamiltonian in Eq.~\eqref{appendix_kagome}.}
\label{fig:Appendix_1}
\end{figure}

Let us consider an extended Bose-Hubbard model on the honeycomb lattice (see Fig.~\ref{fig:squarexmap}(b)) described by the Hamiltonian
\begin{align}
H &= - J \sum_{\mathbf{r},i} \Big(a_{\mathbf{r}+\bs{\delta}_i}^\dagger b^{}_{\mathbf{r}} + \text{H.c.}\Big) + V \sum_{\mathbf{r},i} n^a_{\mathbf{r}+\bs{\delta}_i} n^b_{\mathbf{r}} \nn\\
 &= H_0 + H_V \,,
\end{align}
where the operators $a^{({\dag})}_{\mathbf{r}'}$ and $b^{({\dag})}_{\mathbf{r}}$ annihilate (create) a boson on a $A$ site and a $B$  site, respectively. Consider a site $B$ with lattice coordinate $\mathbf{r}$. Its neighboring $A$ sites are located at $ \mathbf{r} + \bs{\delta}_i$, with $i=1,2,3$ and $\bs{\delta}_1 = \left(a/2, \sqrt{3}a/2\right) $, $\bs{\delta}_2 = \left(a/2, -\sqrt{3}a/2\right)$ and $\bs{\delta}_3 =\left(-a,0\right) $, where $a$ is the lattice spacing, as shown in Fig.~\ref{fig:Appendix_1}(a).

These three bonds define all the links $\ell_i\equiv(\mathbf{r}+\bs{\delta}_i, \mathbf{r})$ where the interaction term is nonvanishing. For simplicity of notation, we have dropped the dependence $\ell_i(\b r)$. However, this dependence will be necessary when considering links belonging to distinct unit cells. We can thus introduce three distinct types of two-particle states 
\be
|d_{\ell_i} \ket \equiv a^\dag_{\mathbf r + \bs{\delta}_i} b^\dag_{\mathbf r} |0\ket = d^\dag_{\ell_i}|0\ket\,.
\ee 
If $V\gg J$, these states are all eigenstates of the interaction Hamiltonian $H_V$ with energy $V_\ell = V$. All the other two-particle states can be written as 
\begin{align}
|\alpha^{(ab)}_{{\mathbf{r}'} \mathbf{r}}\ket &\equiv a^\dag_{{\mathbf r}'} b^\dag_{\mathbf r}|0\ket \,, \,\quad {\mathbf{r}'} \neq \mathbf{r} + \bs{\delta}_i\,, \nn\\
|\alpha^{(aa)}_{{\mathbf{r}'} \mathbf{r}}\ket &\equiv a^\dag_{{\mathbf r}'} a^\dag_{\mathbf r}|0\ket \,, \quad {\mathbf{r}'} \neq \mathbf{r} \,,\nn \\
|\alpha^{(bb)}_{{\mathbf{r}'} \mathbf{r}}\ket &\equiv b^\dag_{{\mathbf r}'} b^\dag_{\mathbf r}|0\ket \,, \,\quad {\mathbf{r}'} \neq \mathbf{r} \,,
\end{align}
and they are eigenstates of $H_V$ with energy $\epsilon_\alpha = 0$. Notice that two-particle states on the same site have been excluded from the Hilbert space because of the hardcore constraint discussed in this work. The separation of energy scales makes possible to use standard perturbation theory \cite{Cohen} to construct an effective Hamiltonian $H_{\mathrm{eff}}$ describing the high-energy manifold of states $|d_\ell \ket$. The matrix elements $\mv{d_{\ell} |  H_{\textrm{eff}} |d_{\ell'}}$ of the effective Hamiltonian are generally written in Eq.~\eqref{eq:perturbth} for the doublon problem with nearest-neighbor interactions, which read for the system discussed in this section as 
\begin{align}
\label{eq:perturbth2}
\mv{d_{\ell} | & H_{\textrm{eff}} |d_{\ell'}} =  V \delta_{\ell, \ell'} + \mv{d_\ell | H_0 | d_{\ell'}} +  \\
 &+ \f 1 2 \sum_{\alpha} \mv{d_{\ell} | H_0 | \alpha} \mv{\alpha | H_0 | d_{\ell'}} \f 2 V \,. \nn
\end{align}

When $\ell = \ell'$, these matrix elements correspond to the \emph{onsite energy} of the effective Hamiltonian, $\epsilon^{\textrm{eff}}_\ell \equiv \mv{d_{\ell} | H_{\textrm{eff}} |d_{\ell}} $. The first term in Eq.~\eqref{eq:perturbth2} is just the binding energy of the doublon. The first order term vanishes in this particular case, and we are therefore left with the calculation of the second order term, namely the matrix elements $\mv{\alpha | H_0 | d_{\ell}} $. Let us consider the link $\ell_3 = (\mathbf r, \mathbf r + \bs\delta_3)$ and the corresponding doublon state $|d_{\mathbf{r}+\bs\delta_3, \mathbf{r}}\ket$. A simple inspection shows that the only states $|\alpha\ket$ that give a nonvanishing contribution are $|\alpha^{(aa)}_{\mathbf{r}+\bs\delta_1, \mathbf{r}+\bs\delta_3}\ket$, $|\alpha^{(aa)}_{\mathbf{r}+\bs\delta_2, \mathbf{r}+\bs\delta_3}\ket$, $|\alpha^{(bb)}_{\mathbf{r}+\bs \delta_3-\bs \delta_1, \mathbf{r}}\ket$ and $|\alpha^{(bb)}_{\mathbf{r}+\bs \delta_3-\bs \delta_2, \mathbf{r}}\ket$, as represented in Fig.~\ref{fig:Appendix_1}(b), and each matrix element is simply equal to $-J$. Summing up all the terms we obtain
\be
\epsilon^{\textrm{eff}}_{\ell_3} = V + \f{4J^2}{V}\,.
\label{appendix_e_eff}
\ee
The calculation for the other two $\ell$ bonds can be performed analogously and yields the same result. 

When $\ell \neq \ell'$, the matrix elements describe \emph{hopping terms} of the effective Hamiltonian, $-J^{\textrm{eff}}_\ell \equiv \mv{d_{\ell} | H_{\textrm{eff}} |d_{\ell}'} $. Let us consider the doublon states at the two bonds $\ell=(\mathbf{r}+\bs{\delta}_3, \mathbf{r})$ and $\ell'=(\mathbf{r}+\bs{\delta}_1, \mathbf{r})$. The first-order term vanishes since the hopping Hamiltonian $H_0$ contains only nearest-neighbor processes. The latter are important for the second-order term. In particular, the intermediate state $|\alpha^{(aa)}_{\mathbf{r}+\bs\delta_1, \mathbf{r}+\bs\delta_3}\ket$ contributes and gives as a result 
\be
-J^{\textrm{eff}}_\ell = J^2/V \,.
\label{appendix_J_eff}
\ee
A simple inspection shows that all possibile effective hopping terms have the same magnitude and generate the kagome effective lattice shown in Fig.~\ref{fig:Appendix_1}(c). As a result, we have just defined all the matrix elements of the effective Hamiltonian $H_\mathrm{eff}$, which we can therefore written in a second-quantized formalism as in Eq.~\eqref{eq:eff}.

It is now useful to clarify the notation in order to work with the effective Hamiltonian in Eq.~\eqref{eq:eff}. To do that, let us make explicit that the effective kagome lattice has a unit cell centered at $\b r$ with three sublattices by writing the doublon operator as $d^\dag_{\ell_i(\b r)} \ra d^\dag_{i}(\b r + \bs \delta_i/2)$, which is shown in Fig.~\ref{fig:Appendix_1}(c). Eq.~\eqref{eq:eff} can therefore be written as 
\begin{align}
 H_\rm{eff} = &- J^\rm{eff} \sum_{\b r, j\neq i} d^\dag_{j}(\b r \! + \!\bs \delta_j/2)\, d^{}_{i}( \b r\! + \!\bs \delta_i/2)  +\nn \\
&- J^\rm{eff} \sum_{\b r, j \neq i} d^\dag_{j}( \b r \!+\! \bs \delta_i \!-\! \bs\delta_j/2)\, d^{}_{i}( \b r\! + \!\bs \delta_i/2) + \nn\\
&+\epsilon^\rm{eff}\sum_{\b r, j}d^\dag_{j}(\b r+ \bs \delta_j/2) d^{}_{j}(\b r+ \bs \delta_j/2)  \,.
\end{align}
It is now possible to define the Fourier transform of the doublon operators as 
\be
d^\dag_{i} (\b r + \bs \delta_i/2) = \f{1}{\sqrt{N_c}} \sum_{\b k} e^{i \b k \cdot (\b r + \bs \delta_i/2)} d^\dag_{i,\b k}\,,
\ee
where $N_c$ is the number of unit cells of the kagome lattice, to obtain the momentum space Hamiltonian
\be
H_\text{eff}(\b k ) =
\begin{pmatrix}
\epsilon_\text{eff} & J_{\b k}^{21} &J_{\b k}^{31} \\[0.5em]
J_{\b k}^{21} & \epsilon_\text{eff} &J_{\b k}^{32} \\[0.5em]
J_{\b k}^{31} & J_{\b k}^{32}& \epsilon_\text{eff} 
\label{appendix_kagome}
\end{pmatrix} \,,
\ee
where $J_{\b k}^{ij} =- 2 J_\text{eff} \cos\left[ \bs k \cdot \left(\frac{\bs \delta_i - \bs\delta_j}{2}\right)\right]$.
Notice that $\b k$ and $\b r$ are the center-of-mass momentum and position, respectively. To further convince oneself of this, it is sufficient to consider the definition of the $d^\dag_{i,\b r}$ operators, namely
 \begin{align}
d^\dag_{i}(\b r + \bs \delta_i/2) &= a^\dag_{\b r+\bs\delta_i} b^\dag_{\b r} = \f{1}{N_c} \sum_{\b k_1,\b k_2} e^{i \b k_1 \cdot (\b r + \bs \delta_i)} e^{i \b k_2 \cdot \b r} a^\dag_{\b k_1} b^\dag_{\b k_2} \nn\\
&= \f{1}{\sqrt{N_c}} \sum_{\b k} e^{i\b k \cdot \b R} d^\dag_{i,\b k}\,,
\end{align}
where $\b R = \b r + \bs \delta_i/2$ is the center-of-mass position, $\b k= \b k_1 + \b k_2$ is the center-of-mass momentum and 
\be
d^\dag_{i,\b k} = \f{1}{\sqrt{N_c}} \sum_{\bs \kappa} e^{i\bs \kappa \cdot \bs \rho} a^{\dag}_{\b k + 2\bs \kappa} b^{\dag}_{\b k - 2\bs \kappa}\,,
\ee
with $\bs \kappa=(\b k _1 - \b k_2)/2$ and $\bs \rho = \bs \delta_i$ the relative momentum and relative position, respectively.

\bibliography{biblio}

\end{document}